\documentclass[transmag]{IEEEtran}
\usepackage{latexsym}
\usepackage{graphicx}
\usepackage{amsfonts,amssymb,amsmath}
\usepackage{algorithm,algorithmic}
\usepackage{hyperref}
\usepackage{color,soul, multirow, makecell}
\usepackage[caption=false, font=footnotesize]{subfig}
\def\BibTeX{{\rm B\kern-.05em{\sc i\kern-.025em b}\kern-.08em T\kern-.1667em\lower.7ex\hbox{E}\kern-.125emX}}
\newcommand\Tstrut{\rule{0pt}{2.6ex}}         % = `top' strut
\newcommand\Bstrut{\rule[-0.9ex]{0pt}{0pt}}   % = `bottom' strut

\begin{document}

\title{Improved {CNN}-based Learning of Interpolation Filters for Low-Complexity Inter Prediction in Video Coding}

\author{Luka Murn, Saverio Blasi, Alan F. Smeaton, \IEEEmembership{Fellow, IEEE}, and Marta Mrak,
\IEEEmembership{Senior Member, IEEE}
\thanks{Manuscript submitted May 18, 2021. The work described in this paper has been conducted within the project JOLT funded by the European Union’s Horizon 2020 research and innovation programme under the Marie Skłodowska Curie grant agreement No 765140.}
\thanks{L. Murn, S. Blasi and M. Mrak are with BBC Research \& Development, The Lighthouse, White City Place, 201 Wood Lane, London,
UK (e-mail: luka.murn@bbc.co.uk, saverio.blasi@bbc.co.uk,
marta.mrak@bbc.co.uk).}
\thanks{A. F. Smeaton is with Dublin City University, Glasnevin,
Dublin 9, Ireland (e-mail: alan.smeaton@dcu.ie)}}

\IEEEtitleabstractindextext{\begin{abstract}The versatility of recent machine learning approaches makes them ideal for improvement of next generation video compression solutions. Unfortunately, these approaches typically bring significant increases in computational complexity and are difficult to interpret into explainable models, affecting their potential for implementation within practical video coding applications. This paper introduces a novel explainable neural network-based inter-prediction scheme, to improve the interpolation of reference samples needed for fractional precision motion compensation. The approach requires a single neural network to be trained from which a full quarter-pixel interpolation filter set is derived, as the network is easily interpretable due to its linear structure. A novel training framework enables each network branch to resemble a specific fractional shift. This practical solution makes it very efficient to use alongside conventional video coding schemes. When implemented in the context of the state-of-the-art Versatile Video Coding (VVC) test model, $0.77\%$, $1.27\%$ and $2.25\%$ BD-rate savings can be achieved on average for lower resolution sequences under the random access, low-delay B and low-delay P configurations, respectively, while the complexity of the learned interpolation schemes is significantly reduced compared to the interpolation with full CNNs.\end{abstract}

\begin{IEEEkeywords}
video compression, motion compensation, interpolation, machine learning, complexity reduction
\end{IEEEkeywords}}

\maketitle

\section{Introduction}

\IEEEPARstart{T}{o} meet the increasing demands for video content at better qualities and higher resolutions, new video compression technology is being researched and developed. The efficiency and versatility of recent Machine Learning (ML) based approaches paved the way for researching more extensive ways to integrate ML solutions into next generation video coding schemes. In this context, the recent MPEG and ITU Versatile Video Coding (VVC) standard \cite{bross2020vvc} already includes some ML-based tools, and it is capable of achieving considerably higher compression efficiency than older video compression technologies.

Deep Convolutional Neural Network (CNN) approaches have been particularly successful in solving tasks such as single image super-resolution \cite{DongLHT15}, image classification, object detection \cite{deepreview}, and colourisation \cite{marccolorizationmmsp}, some fundamental and challenging computer vision problems. However, such approaches are typically very complex, while video compression applications require a careful design of the coding tools, in order to meet the strict complexity requirements posed by practical encoder and decoder implementations. The inference process required by conventional Deep Neural Network (DNN) approaches is inherently complex, limiting their applicability to video coding. As such, learning-based tools within VVC, for instance the Low-Frequency Non-Separable Transform (LFNST) and Matrix Intra-Prediction (MIP) \cite{intradeep}, were carefully designed and simplified throughout the course of the standardisation process. Some other, efficient approaches based on CNNs were proposed to solve video compression tasks such as chroma intra-prediction \cite{marcjournal}, with very promising results.

Additionally, methods based on learning may not generalise to the vast amount of different video content available, especially if the training process is not carefully designed. As such, the algorithms learned by these schemes need to be made transparent and explained, to mitigate potential unexpected outcomes.

\begin{figure}
    \centering
    \includegraphics[width=0.48\textwidth]{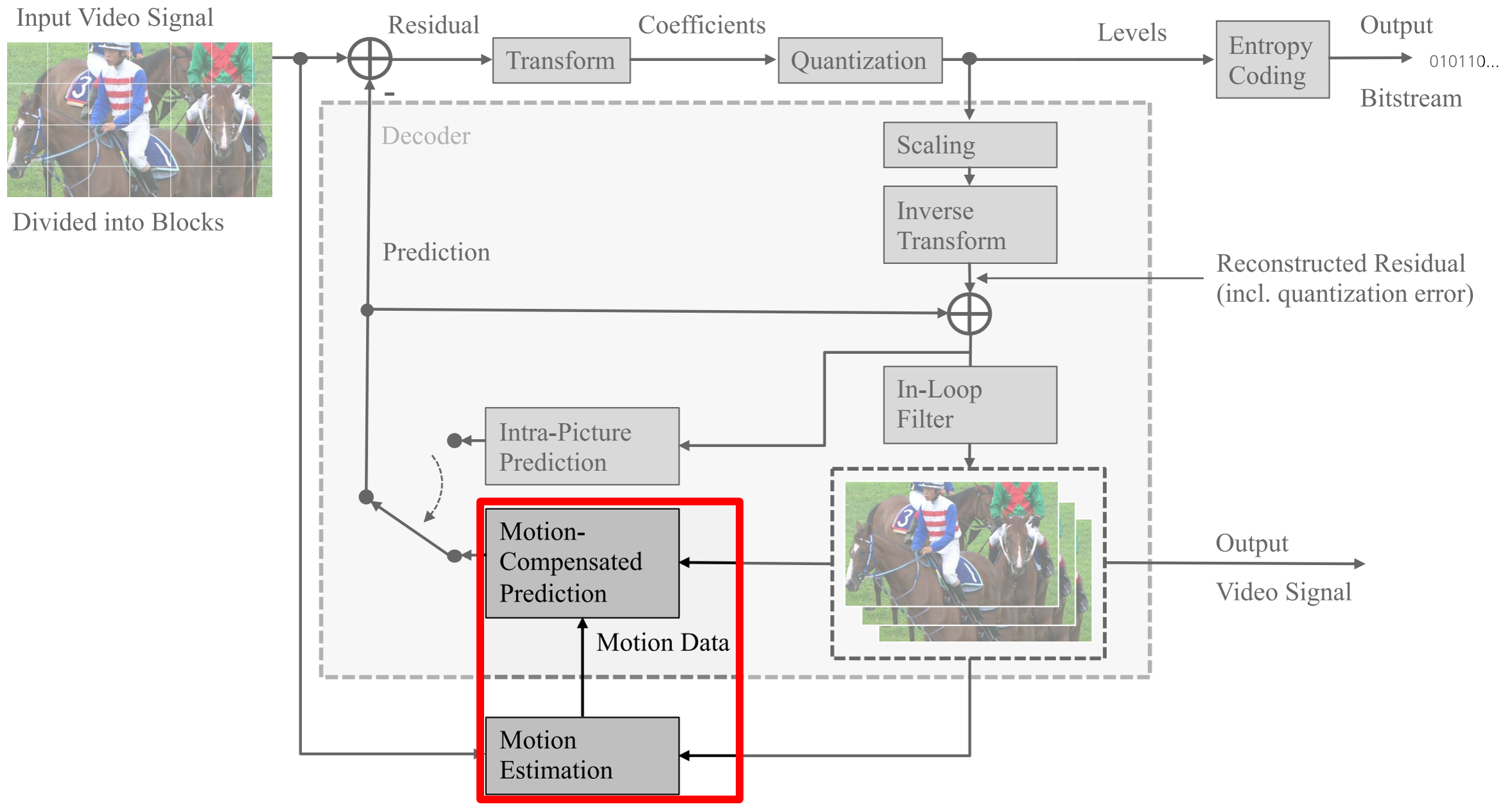}
    \caption{Block diagram of a modern video encoder (e.g. VVC \cite{vvcalgorithms}), with highlighted inter-prediction modules}
    \label{fig:blockdiagram}
\end{figure}

Most video compression technology relies on a block-based hybrid coding scheme, as illustrated in Fig \ref{fig:blockdiagram}. Each frame is split into a number of blocks, using a flexible partitioning process. A block can then be predicted from already reconstructed pixels in either the current frame (intra-prediction) or by means of Motion Compensation (MC) on other reference frames (inter-prediction), using different prediction schemes. The decoder extracts information from the bitstream necessary to perform the correct prediction. Conversely, at the encoder, the different options are usually tested following a Rate-Distortion (RD) strategy. An RD cost is computed, typically a combination of a distortion metric and an estimate of the bitrate required to compress when using such an option. The option with the minimum RD cost is then selected.

Among the various video coding modules, inter-prediction is an essential tool that exploits temporal redundancies among different frames to achieve compression. A Motion Vector (MV) is derived at the encoder side by means of a Motion Estimation (ME) process, and transmitted to the decoder, to identify the spatial displacement between the position of the current block being encoded and the portion of a reference frame that is used for the MC prediction. In order to increase the accuracy of the MC process, sub-pixel interpolation has been widely used in many generations of video coding standards. Interpolation filters are applied to generate sub-pixel locations, which are then used alongside real integer-located pixels to perform MC. Sub-pixel ME simulates frame-to-frame displacements and sub-pixel shifts. Modern standards, such as VVC, may include new, refined interpolation filter sets. For instance, VVC includes new filters for Affine motion  and for Reference Picture Resampling, along with an alternate half-pixel filter \cite{vvcalgorithms}. These developments indicate that the traditional fixed interpolation filters, based on the Discrete Cosine Transform (DCT), may not describe the displacements well enough or capture the diversity within the video data. 

Given the successful application of ML to super-resolution applications, neural networks may be employed within a video coding scheme to produce new, more efficient interpolation filters. In a previous work \cite{murn2020interpreting}, an approach for CNN-based sub-pixel interpolation filters for fractional motion estimation was presented in the context of the VVC standard. This is the first approach that shows gains in the challenging VVC framework, while also reducing the computational requirements to an acceptable level. 
This paper expands on such work \cite{murn2020interpreting}, uncovering new ways to boost the prediction performance of the learned interpolation, without affecting the computational requirements of the prediction process. In particular this paper introduces:
\begin{itemize}
    \item A multi-layer and multi-branch linear convolutional model architecture with shared layers for unified training of the interpolation filter set. The new architecture leverages the commonalities of the interpolation filters in order to train the full set with a single network;
    \item A three-stage training framework enabling competition of network's branches to increase learned prediction performance. Each network branch is trained to resemble a specific quarter-pixel fractional shift, whose weights can then be extracted as interpolation filter coefficients; and 
    \item Detailed experimental analysis of training options and evaluation of the proposed solution in the VVC framework. The evaluation showcases how even a simple network without any non-linear functions and biases can achieve significant coding gains for the fractional interpolation task within video coding.
\end{itemize}

The proposed methodology relies on the powerful learning capabilities of deep CNNs to derive a new quarter-pixel interpolation filter set, even when its non-linear properties are removed to enable a low-complexity inference process required for its implementation within video codecs. The experimental results provide justifications for the proposed designs, and demonstrate that significant compression gains can be achieved for challenging content used for development of video coding standards.

The paper is organised as follows: Section \ref{sec:back} provides a brief background of the related work, Section III describes the proposed methodology and new network architecture, Section IV shows experimental results, with conclusion drawn in Section V.

\section{Background}
\label{sec:back}

Video coding standards typically utilise a set of fixed sub-pixel interpolation filters. In practical solutions, most commonly used interpolation filters are designed for half and quarter pixel positions \cite{thatnokiapaper}. To boost prediction performance, more advanced video coding solutions such as VVC may employ higher sub-pixel precision in certain circumstances, e.g. Affine motion \cite{vvcalgorithms}.

In addition to these approaches where filters were designed based on well known signals, new research has recently emerged where ML is used to derive or refine the sub-pixel samples. In these approaches, the target framework for improvement has been the MPEG/ITU-T High Efficiency Video Coding (HEVC/H.265) standard \cite{hevc2012}. Several methods based on CNNs have been proposed that either improve the input to the interpolation process, by fusing the reference blocks in bi-prediction in a non-linear way \cite{bipredictionfusion, spatiotemporalbipred}, or the output of the process, utilising the neighbouring reconstructed region to refine the prediction \cite{motioncomprefinement}.

In the context of directly deriving the sub-pixel samples, an approach based on super-resolution CNNs to interpolate half-pixel samples was introduced in \cite{Yan2017}, bringing $0.9\%$ Bjøntegaard delta-rate (BD-rate) \cite{bjontegaard2001calculation} bit-rate reductions under low-delay P (LDP) configuration when replacing HEVC luma filters. Furthermore, training separate networks for luma and chroma channels was presented in \cite{ChromaMC}. The resulting models were tested within the HEVC reference software as a switchable interpolation filter, achieving $2.9\%$ BD-rate coding gains under the LDP configuration. As a follow-up to \cite{Yan2017}, a sub-pixel MC was formulated as an inter-picture regression problem rather than an interpolation problem \cite{Yan2019}. The resulting method uses $15$ networks, one for each half and quarter-pixel fractional shift. The input to each network was the decoded reference block for that  position, where the ground truth was the original content of the current block. Different NNs were trained for uni-prediction and bi-prediction and for different Quantisation Parameter (QP) ranges, resulting in a total of $120$ NN-based interpolation filters. Two NN structures were used to train the interpolation filters, where a simpler structure was based on the Super-Resolution CNN (SRCNN) \cite{DongLHT15}, and a more complex model was based on the Variable-filter-size Residue learning CNN (VRCNN) \cite{vrcnn}. When tested on $32$ frames under LDP configuration, $2.9\%$ BD-rate gains were reported for VRCNN and $2.2\%$ for SRCNN with respect to HEVC. The VRCNN implementation had increased the decoding time approximately $220$ times, while the SRCNN increased it nearly $50$ times. In subsequent work, the authors of \cite{Yan2017, Yan2019} have proposed a training scheme driven by invertibility, a marked property of fractional interpolation filters, reaching $4.7\%$ BD-rate reductions for the LDB configuration with a VRCNN architecture, and increasing the decoding time $700$ times on average \cite{InvertibilityFractional}. Lastly, an approach that groups all sub-pixel positions at one level and is trained for different QPs was proposed in \cite{oneforall}, achieving $2.2\%$ coding gains under the LDP configuration.

While these approaches provide powerful prediction performance thanks to NN-generated sub-pixel samples, they have very high computational requirements, increasing the encoding complexity by up to $380$ times \cite{murn2020interpreting}, which makes them inapplicable in real world scenarios. Furthermore, these NN-based interpolation filters are still used as ``black-boxes", i.e. without being able to fully understand how each predicted sample is predicted within the network. As such, many of the produced filters may actually behave very similar to the already existing, fixed interpolation filters. Some of these produced filters may not be useful, unnecessarily over-complicating the video coding scheme. Without complete knowledge on the actual filtering process, it is difficult to compare these filters with the conventional filters, possibly introducing redundancy within the video encoder. Additionally, most of these approaches require GPU-based implementations, which limit their applicability in more conventional CPU-based codec environments.

Recent research in the context of NN-based prediction showed that linear network segments can be deployed in place of more complex schemes with advantages in terms of reducing the complexity of the inference process \cite{Jing2020ImplicitRA}. Application of linear segments has already been presented in the context of intra-prediction for video coding \cite{Santamaria_2020}. Furthermore, in previous work \cite{murn2020interpreting}, a simplified CNN-based filtering process was proposed in the context of inter-prediction in VVC, as opposed to  the approach in \cite{Yan2017, ChromaMC, oneforall} which used HEVC as its baseline encoder. Due to the fact VVC is considerably more efficient than HEVC, it is a much more challenging baseline from which to achieve compression efficiency gains. 

\begin{figure}[ht]
    \centering
    \includegraphics[width=0.48\textwidth]{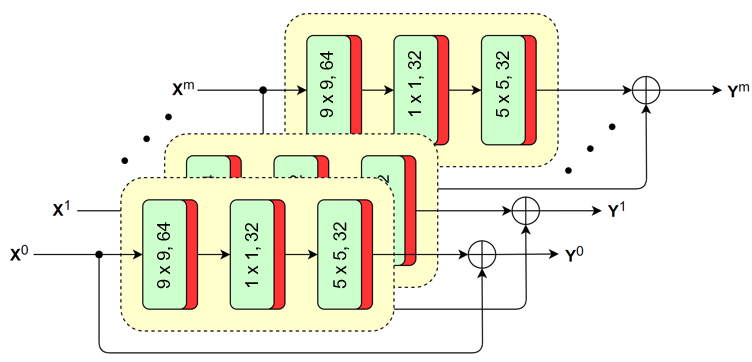}
    \caption{Quarter-pixel interpolation filtering with $ScratchCNN$. Each convolutional kernel size is indicated as (kernel height $\times$ kernel width, number of kernels)}
    \label{fig:scratchcnn15}
\end{figure}

In ScratchCNN \cite{murn2020interpreting}, biases and activation functions were removed from the network architecture, and padding between network layers is not applied in order to reduce the input size from one layer to the next. This resulted in a new, simpler model, with one network trained for each quarter-pixel position, as visualised in Fig.~\ref{fig:scratchcnn15}. ScratchCNN had the ability of collapsing its three-layer structure so only a single matrix multiplication would need to be used at inference time, meaning that filter coefficients could be directly extracted from the learned network weights. These new, alternative learned filter coefficients could then be analysed easily to understand what exactly had been learned, rather than blindly implementing a full network with several layers. Moreover, the full, three-layer network with its large set of weights is no longer needed. The prediction process only requires the new, much smaller number of coefficients. Such an approach significantly decreased the storage and complexity requirements of its implementation, and also allowed tests to run on conventional, CPU-based computing architectures.

However, the existing approach still requires a significant number of networks to be trained, one for each fractional pixel location. Additionally, the training process could be improved to better utilise a neural network's learning capabilities. This paper proposes to solve these issues. The ScratchCNN model \cite{murn2020interpreting} in Fig.~\ref{fig:scratchcnn15} was used as basis for this approach, as it is the only model with acceptable complexity.

\section{Methodology}
This section introduces the general network architecture of the proposed approach and the novel methodology employed to produce the learned filters, as described in detail in Section \ref{subsec:genarch}. The framework enables the training of filters with shared layers, referred to as SharedCNN (Section \ref{subsec:sharedcnn}), and in competition with VVC filters, referred to as CompetitionCNN (Section \ref{subsec:compcnn}). Additionally, Section \ref{subsec:traincompcnn} explains the multi-stage training scheme which ensures that the learned interpolation filter set of CompetitionCNN is unlike the already present VVC filter set. In Section \ref{subsec:learnedcoeffs}, details are provided on how the learned predictors can be simplified into interpolation filters similar to those in \cite{murn2020interpreting}. The produced filters can be easily integrated into the VVC framework as switchable options. This means that for a given sub-pixel MV, the encoder will test both the conventional filters as well as the proposed, trained filters, and select the best option, which is signalled to the decoder in the bitstream.

\subsection{General network architecture}
\label{subsec:genarch}

The learned filters can be used to derive up to quarter-pixel precision predictions. Each pixel therefore requires $15$ fractional samples to be produced, to account for all quarter and half pixel locations between a given integer sample, and its neighbouring integer samples on the bottom and on the right. For a given reference block, or input $\mathbf{X}^m$, the ScratchCNN network \cite{murn2020interpreting} as in Fig.~\ref{fig:scratchcnn15} predicts residuals $\mathbf{R}^{m}$, where $m=0,...,14$ corresponds to each of 15 half and quarter pixel positions. Each residual is then added to the input reference samples to form the prediction $\mathbf{Y}^{m}$ representing values at a quarter-pixel position as:

\begin{equation}
\label{eq:network}
\mathbf{Y}^{m} = \mathbf{R}^{m} + \mathbf{X}^m.
\end{equation}

\begin{figure}
    \centering
    \includegraphics[width=0.3718\textwidth]{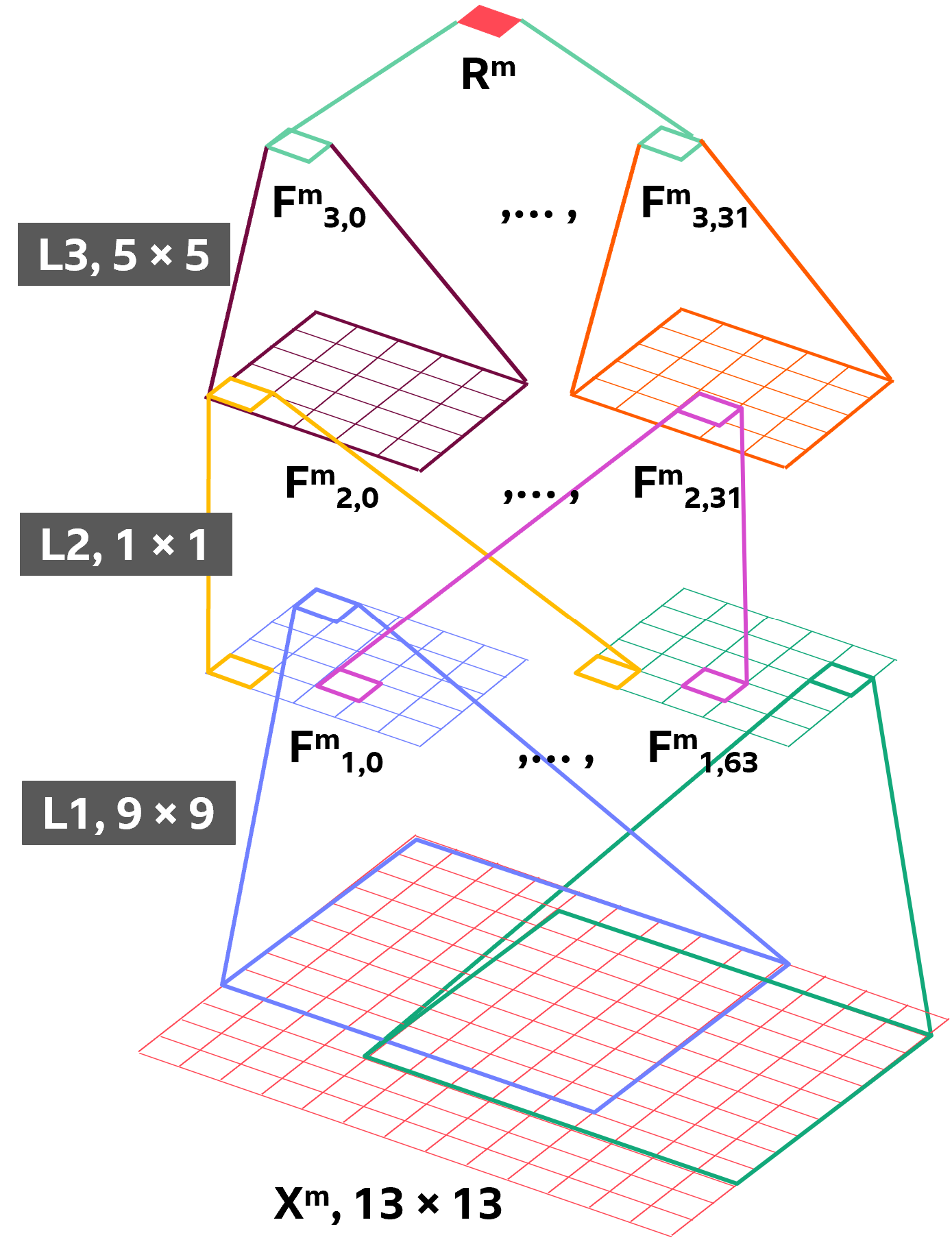}
    \caption{Fractional pixel derivation process for branch $m$ of an NN interpolation filter set. The proposed approach requires $13\times13$ input samples.}
    \label{fig:interpret}
\end{figure}

To obtain $\mathbf{Y}^{m}$, the central pixel from the $13 \times 13$ window of $\mathbf{X}^m$ is used in Eq. \ref {eq:network}. The proposed model is a three-layer convolutional network, which contains $64$ $9\times9$ convolutional kernels in the first layer, followed by $32$ $1\times1$ kernels in the second layer, while the last layer consists of $32$ $5\times5$ convolutional kernels. This structure is based on the well-known SRCNN model \cite{DongLHT15}. However, it is modified by removing activation functions and biases. Additionally, padding between layers is not applied in order to use only the available input samples. Thus, the model can be expressed as a series of linear operations from the input to the output. The process of acquiring a single pixel value from the input area is illustrated in Fig.~\ref{fig:interpret}. Due to the convolutional kernel dimensions ($9\times9$, $1\times1$, $5\times5$), an input area of $13\times13$ is required to compute an interpolated output sample.

The approach, as described in \cite{murn2020interpreting}, requires separate networks for each of the interpolated pixel samples. As illustrated in Fig.~\ref{fig:interpret}, the first layer, L$1$, extracts features from regions of input area $\textbf{X}^m$ by applying $9\times9$ convolutional kernels, denoted as $\mathbf{K}^m_{1,i}$, to obtain the first layer output, $\mathbf{F}^m_1$. It is defined as:

\begin{equation}
\label{eq:firstlayer}
\mathbf{F}^m_{1,i} = \mathbf{K}^m_{1,i} * \mathbf{X}^m,
\end{equation}

\noindent where $i=0,...,63$. The second layer, L$2$, combines features extracted from the same regions to retrieve its output, $\mathbf{F}^m_2$, as:
\begin{equation}
\label{eq:secondlayer}
%\mathbf{F}^m_{2,j} = \displaystyle\sum_{i=0}^{63} %\mathbf{K}^m_{2,j} \mathbf{F}^m_{1,i},
\mathbf{F}^m_{2,j} = \displaystyle\sum_{i=0}^{63} K^m_{2,j} \mathbf{F}^m_{1,i},
\end{equation}

\noindent where $K^m_{2,j}$ corresponds to $1\times1$ convolutional kernels and $j=0,...,31$. The network then combines all extracted features to reconstruct the pixel $\mathbf{R}^m$ from $\mathbf{F}^m_3$ feature maps comprising of $5\times5$ convolutional kernels $\mathbf{K}^{m}_{3,j}$ as:

\begin{equation}
\label{eq:thirdlayer}
\mathbf{F}^{m}_{3,j} = \mathbf{K}^{m}_{3,j} * \mathbf{F}^m_{2,j}, 
\end{equation}

\noindent and their summation for each $j=0,...,31$ as:

\begin{equation}
\label{eq:output}
\mathbf{R}^m = \displaystyle\sum_{j=0}^{31} \mathbf{F}^{m}_{3,j}.
\end{equation}

Lastly, the required, non-separable $2$D interpolation filter set $\mathbf{M}^m$ can be obtained from the series of convolutional kernel weights $\mathbf{K}^m_{1,2,3}$ which were multiplied with values in the input $\mathbf{X}^m$. The interpolated sample can be described as:

\begin{equation}
    \label{eq:interpret}
\mathbf{R}^m = \mathbf{M}^m * \mathbf{X}^m = \mathcal{F}(\mathbf{K}^m_3, K^m_2, \mathbf{K}^m_1, \mathbf{X}^m).
\end{equation}

\noindent The coefficients of each filter $\mathbf{M}^m$ represent the contribution of reference samples towards the final prediction in a $13\times13$ surrounding area. The filter set corresponds to the $15$ possible quarter-pixel fractional positions, each trained with 15 ($m=0,...,14$) separate sets of samples $\mathbf{X}^m$ already classified before training.

\subsection{SharedCNN}
\label{subsec:sharedcnn}

\begin{figure}
    \centering
    \includegraphics[width=0.48\textwidth]{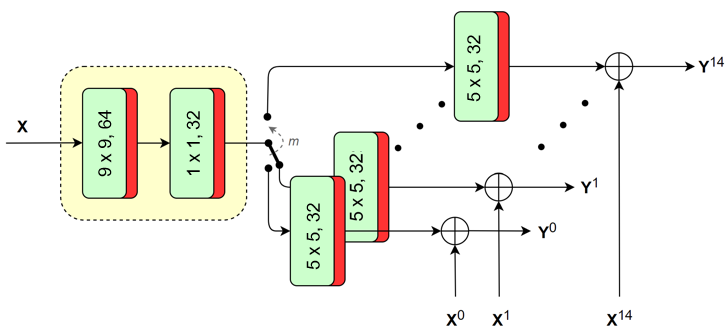}
    \caption{$SharedCNN$ architecture with a trunk with shared layers}
    \label{fig:sharedcnn}
\end{figure}

The ScratchCNN architecture requires completely independent networks to be trained for each of the 15 sub-pixel positions. This is an inadequate solution. The networks are not flexible, as they do not share any information amongst themselves, while at the same time trying to approximate a very similar problem. The training process is computationally expensive,  requiring to retrain each network whenever there is a need to expand or modify the training set. For that reason, a new architecture called SharedCNN is proposed where a so-called trunk, comprising of the first two layers in the network, is shared among all branches which still correspond to individual sub-pixel positions. This is illustrated in Fig.~\ref{fig:sharedcnn} where the branches are trained for only one sub-pixel position, and the trunk is trained for all positions. Thanks to such an architecture, the networks require considerably fewer coefficients to be trained and updated, which can improve learning performance.

Both the ScratchCNN linear model, as well as the new shared model architecture SharedCNN in Fig.~\ref{fig:sharedcnn}, require labelled samples for training so that specific ScratchCNN or specific branch in the network with shared trunk can be updated for each fractional shift $m$ at a time. The equations describing the SharedCNN network are equivalent to Eq. \ref {eq:network} - \ref {eq:interpret}, with specifically defined convolutional layers 1 and 2 which are the same for all sub-pixel positions:

\begin{equation}
    \label{eq:shared}
\mathbf{K}_{1} = \mathbf{K}^{m}_{1}, 
K_{2} = K^{m}_{2}, 
\forall m=0,...,14.
\end{equation}

From the inference perspective, shared layers are used with each individual branch to produce a non-separable $2$D filter for each fractional location, as in ScratchCNN.

\subsection{CompetitionCNN}
\label{subsec:compcnn}

\begin{figure}
    \centering
    \includegraphics[width=0.48\textwidth]{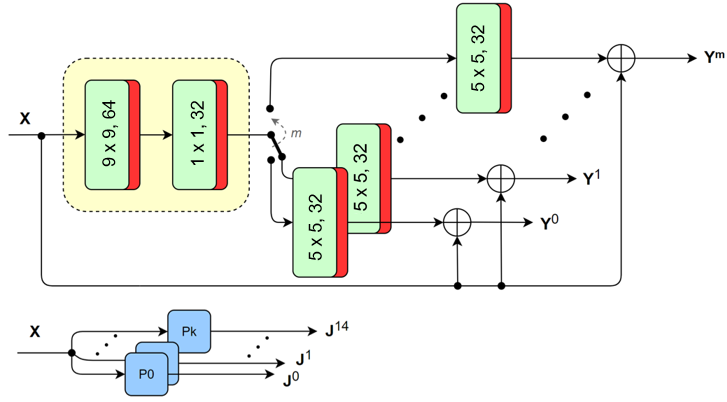}
    \caption{$CompetitionCNN$ network architecture}
    \label{fig:competitioncnn}
\end{figure}

All frameworks discussed in this section so far rely on training with labelled data which forces interpolation filters to be specialised for specific pre-defined sub-pixel positions. However, such labelling may not be optimal and may limit adaptability of learned filters. Furthermore, learned interpolation filters are intended to be integrated into a video codec which can select between the standard filters and learned filters on a block basis. This means that during training it may be better to prevent back-propagation for some blocks which are better predicted with standard filters. 
In order to mitigate these problems, a new framework can be defined, on top of the architecture with shared layers in Fig.~\ref{fig:sharedcnn}, by evaluating the loss of each prediction from both all neural network filters, as well as standard filters ($P_0$ to $P_{14}$). The network is named CompetitionCNN and is detailed in Fig.~\ref{fig:competitioncnn}. The model contains two shared layers, with the last layer consisting of $b=0,...,14$ branches which, during training, produce predictions for every input block (i.e. not only for pre-defined sub-pixel positions). For the inference stage, these branches will be used to create 15 different filters $\mathbf{M}^m$. The process of acquiring an interpolated sample is equivalent to the one in SharedCNN and it corresponds to:

\begin{equation}
    \label{eq:compinterpret}
\mathbf{R}^m = \mathbf{M}^m * \mathbf{X} = \mathcal{F}(\mathbf{K}^m_3, K_2, \mathbf{K}_1, \mathbf{X}).
\end{equation}

The input-output pairs for training the proposed approaches are formed by VVC. The ground truth is the uncompressed rectangular block in the current frame, while the input to the network is the decoded reference block that was chosen for sub-pixel interpolation. An important aspect of the generated pairs that needs to be preserved is the sub-pixel shift $m$ from the reference input block to the output. The resulting dataset $D$ contains subsets $d^m$ and their relationship is described as:

\begin{equation}
\label{eq:subdataset}
D = \bigcup d^m.
\end{equation}

Due to its architecture constraint, each ScratchCNN model as well as SharedCNN had to be trained with a specific subset $d^m$, consisting of pairs $(\mathbf{X}^m, \mathbf{Y}^m)$. However, the CompetitionCNN relaxes this fragmentation, as explained in the next subsection, using the whole dataset $D$ for all branches.

\subsection{Three-stage training for CompetitionCNN}
\label{subsec:traincompcnn}

In general, the network architecture of SharedCNN and CompetitionCNN are the same. However, their main difference is showcased in their respective training frameworks. To ensure that each branch of the final layer in the proposed CompetitionCNN competes with others when trained on dataset $D$, a three-stage training framework is proposed. The framework is inspired by a similar approach when training unique intra-prediction modes based on neural networks for its inclusion in the VVC standard \cite{intradeep}.

The loss function used for training all neural networks in this paper is the Sum of Absolute Differences (SAD) loss. SAD is also used in CompetitionCNN to decide which branch will be used for back-propagation. For each branch $b$ of the CompetitionCNN network, the loss $l^b$ is calculated as:

\begin{equation}
    \label{eq:lossbranch}
l^b = SAD(\mathbf{Y}^b, \mathbf{GT}) = E[|\mathbf{Y}^{b} - \mathbf{GT}|],
\end{equation}

\noindent where $\mathbf{GT}$ denotes the ground truth, i.e. the original uncompressed block which needs to be predicted. The three stages of the proposed training are performed according to the following description.

$Stage \; 1$: To set the weights of the shared trunk and to also avoid the issue of branch starvation, where only a few branches in the final layer of the network are back-propagated while other $\mathbf{K}^m_3$ kernels are mostly ignored due to their initial weight setting, the training loss $L_1$ of the first training epoch jointly updates all layers of the network with the loss of the whole dataset $D$ in all branches, as follows: 

\begin{equation}
\label{eq:lossepoch1}
L_1 = \displaystyle\sum_{b} l^b_D.
\end{equation}

$Stage \;2$: The  critical  requirement  of  the  proposed  architecture  is the emergence of $15$ unique interpolation filters, that mostly correspond to particular quarter-pixel fractional shifts. During the second epoch, each branch $b$ of the last layer of the neural network is pre-trained with a separate subset $d^m$ of the dataset, such that $m = b$ , equivalent to each quarter-pixel position. The total loss $L_2$ is defined as:

\begin{equation}
\label{eq:lossepoch2}
L_2 = \displaystyle\sum_{b, m=b} l^b_{d^m},
\end{equation}

\noindent where only $b_{d^m}$, with $m=b$, is used to update each branch. By applying this step, a higher correlation between a branch output and a distinct quarter-pixel filter is established to further prevent starvation. Earlier experiments without this step resulted in several filters being a combination of similar filters. 

\begin{figure*}[t!]
    \centering
    \includegraphics[width=\textwidth]{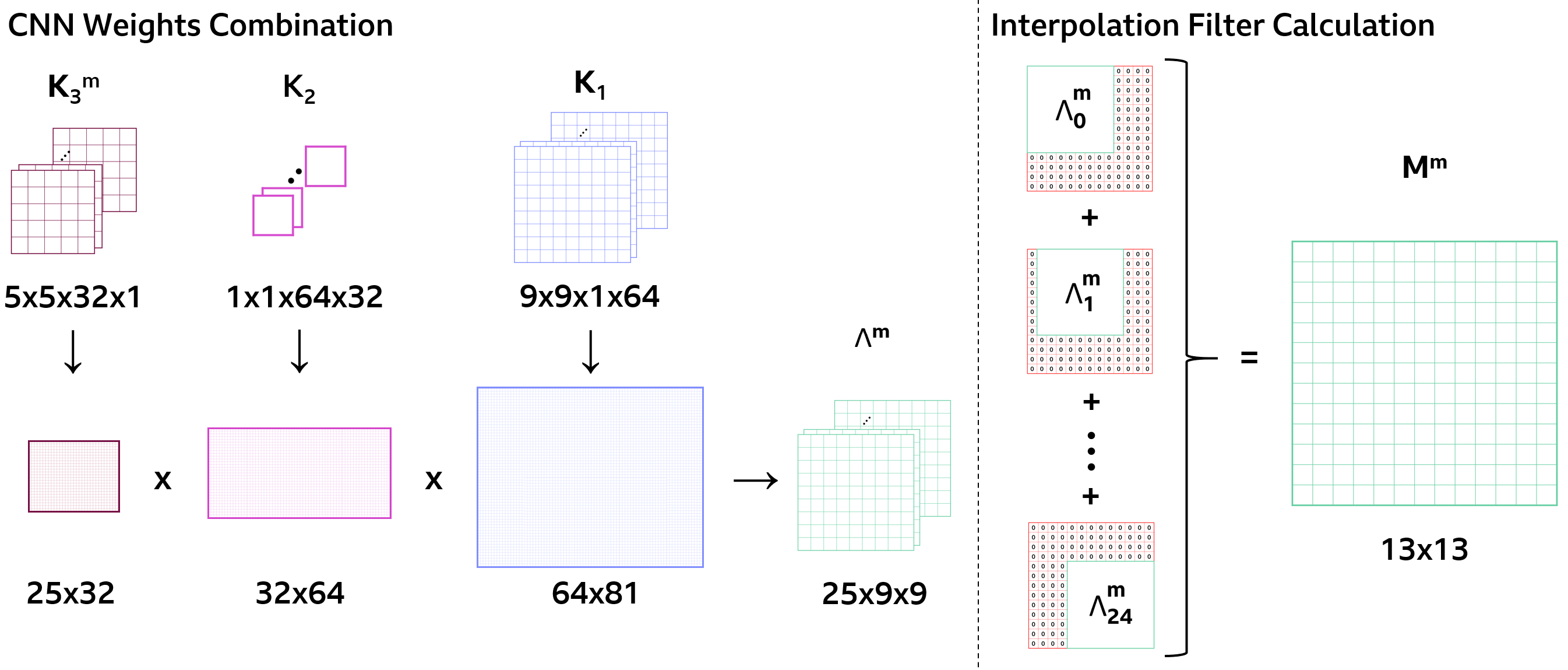}
    \caption{A visualisation of the interpolation filter extraction process. Weights from all layers are combined into $\mathbf{\Lambda}^m$. Each $9\times9$ coefficient matrix is placed within an $13\times13$ zero array before being summed up to obtain $\mathbf{M}^m$ }
    \label{fig:filterextraction}
\end{figure*}

$Stage \; 3$: Lastly, after the two pre-training steps are finished, training is commenced with a loss $L_3$ that updates the two shared layers and only the branch of the network which minimizes the loss for particular blocks, if the prediction it produces has lower SAD than any of the standard predictions $\mathbf{J}^n$, $n=0,...,14$, obtained by corresponding standard filters which are denoted as $P_n$, Fig.~\ref{fig:competitioncnn}. Index $\mu$ of the network's branch that produces minimal loss for given block from $D$ is found as:

\begin{equation}
    \label{eq:mu}
    \mu = \operatorname*{argmin}_b l^b_D.
\end{equation}

\noindent If a network filter fares better than all $\mathbf{J}^n$, i.e. if

\begin{equation}
\label{eq:backpropfull}
l^\mu < \operatorname*{min}_n SAD(\mathbf{J}^n, \mathbf{GT}), 
\end{equation}

\noindent then back-propagation in $\mu^{th}$ branch is commenced for a given block. 

This stage ensures that the network competes with the traditional VVC filters during the network training process already, leading to a closer representation of its final implementation form – a switchable interpolation filter set. 

\subsection{Learned interpolation filter coefficients}
\label{subsec:learnedcoeffs}
After training a network with shared layers, whether it is SharedCNN or CompetitionCNN, each $2$D interpolation filter of size $13\times13$ can be constructed from the network's trunk combined with each branch using Eq.~\ref{eq:compinterpret}, by following the principle visualised in Fig.~\ref{fig:interpret}. 

In Fig.~\ref{fig:filterextraction}, this process is explained in more detail. Since the architecture of a network with shared layers does not contain any non-linear functions or biases, a master kernel matrix $\mathbf{\Lambda}^m$ can be formulated as a combination of network weights $\mathbf{K}^m_3$, $K_2$, $\mathbf{K}_1$ for each branch $m=0,...,14$ of the network. As illustrated in Fig.~\ref{fig:interpret}, the $\mathbf{\Lambda}^m$ matrix is applied onto $9\times9$ patches of an input $\mathbf{X}^m$. By placing each of the $25$ $9\times9$ matrices into the correct position within a $13\times13$ array and summing them all up, the interpolation filter $\mathbf{M}^m$ is created from a trained CNN. Note that the networks described in this paper learn residuals, so a value of 1 needs to be added to the central position of the learned interpolation filter $\mathbf{M}^m$ before using it in a video codec environment.

\begin{figure*}
    \centering
    \includegraphics[width=\textwidth]{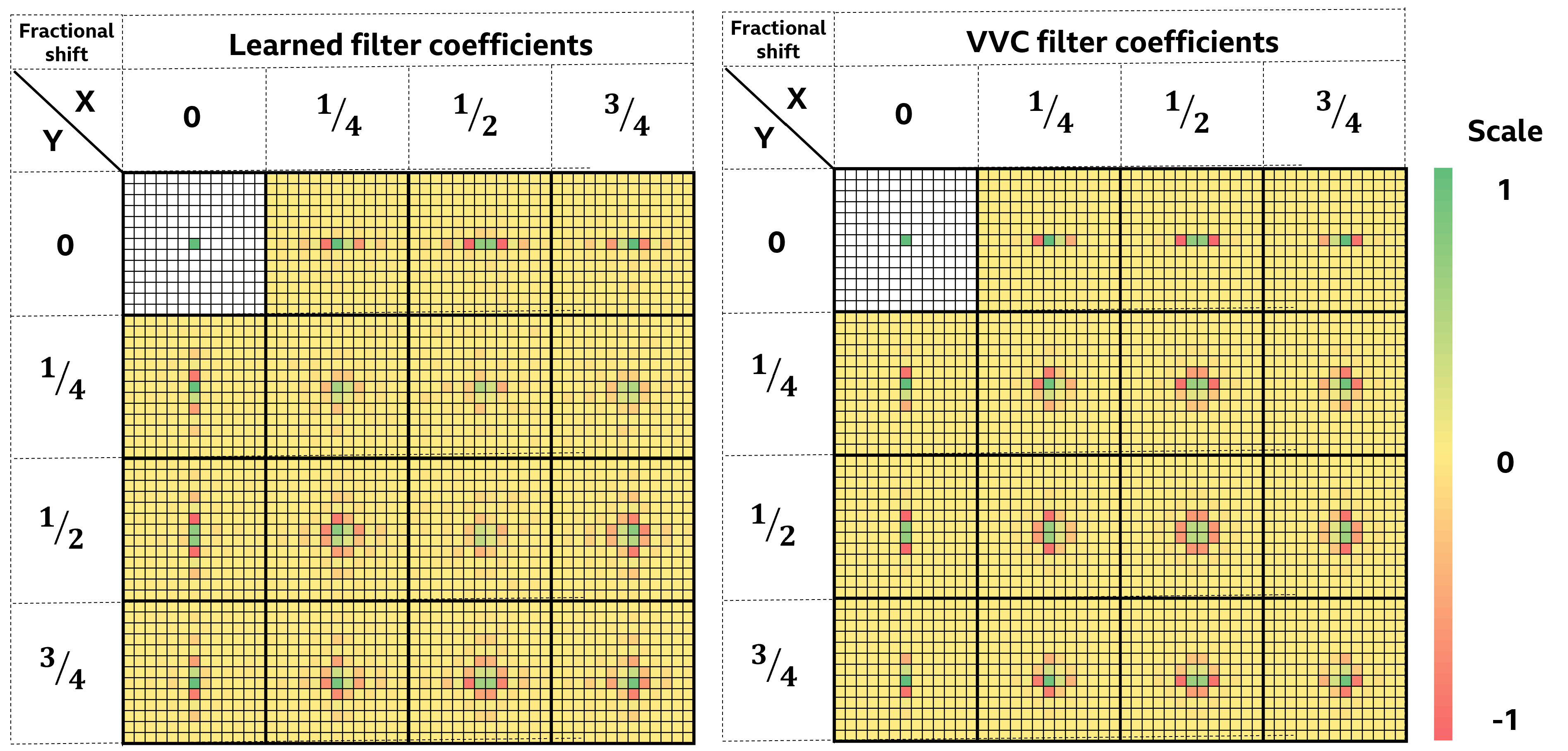}
    \caption{A comparison of $15$ learned $13\times13$ interpolation filters, one for each quarter-pixel position, and ones already present in VVC. For the $(0, 0)$ position, only the central pixel is considered}
    \label{fig:filters}
\end{figure*}

The filter coefficients corresponding to each sub-pixel shift, for CompetitionCNN, are visualized in Fig.~\ref{fig:filters}. It can be observed how the learned coefficients are correlated to the VVC filter coefficients. Only a portion of these coefficients actually contribute to the final pixel prediction, implying that a handful of them can be discarded with a negligible decline in filter efficiency. For example, a list of significant coefficients can be extracted for each filter $\mathbf{M}^m$. Additionally, extracted filter coefficients can also be quantised to integer values to resemble filter sets already present in the VVC standard, which are whole numbers $\mathbb{Z}$ in the range $[-2^6, 2^6]$. In this paper, filter coefficients are integrated within VVC in floating point precision, in the $[-1, 1]$ range. Any simplification approaches would need to be highly optimised to further reduce coding complexity without a significant decline in coding performance. This is left for future work as such optimisations would have to be harmonised with the design of other ML tools in a video codec.

The learned interpolation filter set is implemented into VVC's test model as switchable interpolation filters. For every potential fractionally interpolated block, both the existing filters and new, learned filters are considered before deciding the preferred choice. The selection is handled at a Coding Unit (CU) level, indicated as a flag in the compressed bitstream. In merge mode, blocks to be merged inherit the flag and consequently the choice of the interpolation filter. 

\section{Experimental results}

In this section, a description of the dataset, training and testing configuration is presented, followed by experiments which validate a linear network approach for sub-pixel interpolation filtering\footnote{An open-source software package with the full implementation of this paper is available at \url{https://github.com/bbc/cnn-fractional-motion-compensation}}. Next, prior and proposed linearised network architectures are compared, before a final test is performed on Common Test Condition (CTC) video sequences, as defined by JVET \cite{boyce2018ctc}, with several encoder configurations. CTC video sequence classes are defined as sequences of the same spatial resolution. 

\begin{table*}[t]
\caption{Loss comparison of linear and non-linear network structures, on Class D sequences}
\centering
\label{tab:netstructures}
\begin{tabular}{c|c|c|c|c|c} 
\hline
\hline
	Model & Convolutional kernels & Number of networks & Activation functions & Biases & SAD\Tstrut\Bstrut\\
\hline
\multirow{5}{*}{\makecell{ScratchCNN}} & \multirow{4}{*}{\makecell{L1: 9x9,64\\L2: 1x1,32\\L3: 5x5,32}} & \multirow{5}{*}{\makecell{15}} & Yes & Yes & 21.74\Tstrut\Bstrut\\
\cline{4-6}
& & & Yes & No & 21.75\Tstrut\Bstrut\\
\cline{4-6}
& & & No & Yes & 21.72\Tstrut\Bstrut\\
\cline{4-6}
& & & No & No & 21.68\Tstrut\Bstrut\\
\cline{2-2}\cline{4-6}
& L1: 13x13,1 & & No & No & 21.75\Tstrut\Bstrut\\
\hline
\hline
\end{tabular}
\end{table*}

\subsection{Dataset generation}
\label{sub:data}

The training dataset for the proposed approach of fractional interpolation using CNNs is created by encoding the video sequence \textit{BlowingBubbles} within VVC Test Model (VTM) 6.0 \cite{chen2019algorithmvtm6}, under the Random Access (RA) configuration.

In order to generate accurate training data, a set of restrictions has been applied to the CTC conditions. The flags are: Triangle=0, Affine=0, DMVR=0, BIO=0, WeightedPredP=0, WeightedPredB=0. Affine, DMVR and Triangle modes of fractional interpolation have been disabled to generate a larger quantity of data from a selected sequence. Weighted prediction is turned off as it may contribute to a less accurate selection of sub-pixel positions $m$ when blocks with lower weights are considered. To avoid training different networks for different Quantisation Parameters (QP), it is set to $27$. Previous experiments have shown how filters generated from such a dataset generalize well when tested on sequences coded with various QPs. There is no restriction on block sizes, the network is trained on all rectangular block examples. As NN padding is avoided by the introduction of larger receptive fields, each convolutional operation reduces the block size of an example. Thus, the input reference block needs to be larger than the output prediction, comprising a $13\times13$ surrounding area around the required predicted pixels. If the input block is on the boundary of the frame, repetitive padding is applied to add the required pixels, a typical approach used in many MPEG standards \cite{mpeg4}.

After extracting the luma component data from VVC, the dataset is balanced before network training in terms of having equal number of examples for each extracted fractional shift and block size. As the \textit{BlowingBubbles} sequence is of $416\times240$ spatial resolution, the dataset contains block sizes up to $32\times32$, as larger block partitioning is very scarce for the selected video input.

\subsection{Training and testing configuration}

The neural network was trained for $1000$ epochs with a batch size of $32$, using an Adam optimizer and a learning rate of $0.0001$. Early stopping is enabled if the loss does not reduce for more than $50$ epochs, i.e. when it converges. To avoid possible large gaps in loss optimization between the pre-training and training phases, gradient clipping by norm is implemented within the network, with the norm value set to $5.0$.

The switchable filter implementation was tested in a modified VTM 6.0 encoding environment. The inter-prediction encoding conditions were simplified to resemble those of HEVC, with a number of restrictions imposed to coding tools and algorithms. The same encoding configuration flags were used as for generating the dataset, along with these additional restrictions: MHIntra=0, SBT=0, MMVD=0, SMVD=0, IMV=0, SubPuMvp=0, TMVPMode=0.

\subsection{Using linear, multi-layer networks}
To confirm that adopting a linear network approach without biases for creating fractional interpolation filters benefits the prediction performance, a comparison of different network architectures was undertaken. A test was performed on Class D CTC video sequences, comparing SAD losses on fractionally interpolated blocks with ScratchCNN, SharedCNN and CompetitionCNN after training. All input blocks were interpolated by the described networks and then compared to the ground truth block, as defined in Eq.~\ref{eq:lossbranch}. This loss was then compared with the loss of a block interpolated using VVC filters, with the minimum value of those two losses used as the final error. Using the minimum SAD loss between the network filters and VVC filters mimics the behavior of a switchable filter implementation in an actual video codec.

From Table \ref{tab:netstructures}, it can be observed that the difference in performance between all three-layer ScratchCNN network architectures is negligible. Furthermore, adding biases and/or activation functions after the first and second layer of the network does not contribute to its performance. Consequently, removing these parameters creates a linear network structure that enables $2$D interpolation filters to be extracted from the trained architecture, explained in Eq.~\ref{eq:interpret}, which leads to a much more streamlined and low-complexity implementation within a video codec.

Additionally, another test was performed to demonstrate how training a multi-layer linear network leads to a closer representation of the underlying solution than training a simple one-layer network. After training, a multi-layer network collapses into a single layer, behaving the same as a one-layer network, but the learned coefficients are different. In Table \ref{tab:netstructures}, it can be observed how the network trained with only one $13\times13$ convolutional kernel has higher SAD loss than other ScratchCNN structures.

To demonstrate how this loss difference is reflected within VVC, interpolation filters extracted from both the three-layer and one-layer ScratchCNN were integrated as switchable filters in the codec. In Table \ref{tab:multilinear}, the two linear structures are compared, with the multi-layer architecture displaying notably better performance across the RA, low-delay B (LDB) and low-delay P (LDP) encoding configurations. In the case of RA, the coding gains for the luma component, as expressed in the BD-rate, more than double, going from $-0.10\%$ to $-0.26\%$. The BD-rate metric expresses the objective video quality when encoding content at different bit rates. Negative values indicate better compression performance for the same objective quality, compared to the baseline.

\begin{table}[t!]
\caption{BD-rate luma comparison of one-layer and multi-layer network structures, on 32 frames for Class D sequences}
\centering
\label{tab:multilinear}
\begin{tabular}{c|c|ccccc} 
\hline
\hline
\multirow{2}{*}{\makecell{Model}} & \multirow{2}{*}{\makecell{Convolutional\\kernels}}                     & \multicolumn{5}{c}{Encoder configuration}\\
\cline{3-7}
&& \multicolumn{1}{c}{RA} && \multicolumn{1}{c}{LDB} && \multicolumn{1}{c}{LDP}\Tstrut\Bstrut\\
\hline
\multirow{2}{*}{\makecell{ScratchCNN}} & {\makecell{L1: 9x9,64\\L2: 1x1,32\\L3: 5x5,32}} & -0.26\% && -0.58\% &&	-1.11\%\Tstrut\Bstrut\\
\cline{2-7}
& L1: 13x13,1 & -0.10\% && -0.37\% &&	-1.00\%\Tstrut\Bstrut\\
\hline
\multirow{2}{*}{SharedCNN} & \multirow{4}{*}{\makecell{L1: 9x9,64\\L2: 1x1,32\\L3: 5x5,32}} & \multirow{2}{*}{-0.34\%} && \multirow{2}{*}{-0.53\%} && \multirow{2}{*}{-1.21\%}\\
& & & & & \\
\cline{1-1}\cline{3-7}
\multirow{2}{*}{CompetitionCNN} && \multirow{2}{*}{-0.62\%} && \multirow{2}{*}{-1.06\%} && \multirow{2}{*}{-2.21\%} \\
& & & & & \\
\hline
\hline
\end{tabular}
\end{table}

\begin{table*}[ht!]
\footnotesize
\caption{Coding performance of the proposed approach for RA, LDB and LDP configurations;\\ BD-rate for luma; tested on CTC and non-CTC sequences.}
\label{tab:bdr_configs}
\centering
\begin{tabular}{c|ccccc|ccccc}
\hline
\hline
\multirow{2}{*}{\makecell{\textit{Sequence}\\\textbf{Set / Class}}} & \multicolumn{5}{c|}{ScratchCNN \cite{murn2020interpreting}} & \multicolumn{5}{c}{CompetitionCNN (proposed)}\\
\cline{2-11}
& RA && LDB && LDP & RA && LDB && LDP \Tstrut\Bstrut\\
\hline
\textit{BasketballDrill} \textbf{(C)} & -0.15\% && 0.11\% &&	-0.28\% & -0.33\% && -0.79\% && -1.38\% \Tstrut\Bstrut\\
\textit{BQMall} \textbf{(C)} & -0.32\% && -0.69\% &&	-1.25\% & -0.66\% && -0.87\% &&	-1.79\% \Tstrut\Bstrut\\	
\textit{PartyScene} \textbf{(C)}	& -0.82\% && -1.92\% && -3.22\% & -1.32\% && -2.18\% && -3.78\% \Tstrut\Bstrut\\
\textit{RaceHorses} \textbf{(C)} & 0.14\% && 0.19\% && 0.19\% & -0.62\% && -0.43\% && -0.73\% \Tstrut\Bstrut\\	
\hline
\textbf{CTC ClassC Overall} & \textbf{-0.29\%} && \textbf{-0.58\%} && \textbf{-1.14\%} & \textbf{-0.73\%} && \textbf{-1.07\%} && \textbf{-1.92\%} \Tstrut\Bstrut\\
\hline
\textit{BasketballPass} \textbf{(D)} & -0.14\% && -0.33\% &&	-0.52\% & -0.43\% && -0.61\% &&	-1.32\% \Tstrut\Bstrut\\	
\textit{BQSquare} \textbf{(D)} & -1.35\% && -3.02\% && -4.54\% & -1.42\% && -3.06\% && -5.54\% \Tstrut\Bstrut\\
\textit{BlowingBubbles} \textbf{(D)} & -0.90\% && -2.18\% &&	-3.14\% & -1.16\% && -2.02\% &&	-2.96\% \Tstrut\Bstrut\\	
\textit{RaceHorses} \textbf{(D)} & 0.04\% && 0.21\% && 0.02\% & -0.17\% && -0.23\% && -0.49\% \Tstrut\Bstrut\\
\hline
\textbf{CTC ClassD Overall} & \textbf{-0.59\%} && \textbf{-1.33\%}	&& \textbf{-2.04\%} \Tstrut\Bstrut & \textbf{-0.80\%} &&	\textbf{-1.48\%} && \textbf{-2.58\%}  \Tstrut\Bstrut\\
\Xhline{3\arrayrulewidth}
\textbf{CTC ClassC\&D Overall} & \textbf{-0.44\%} && \textbf{-0.95\%}	&& \textbf{-1.59\%} \Tstrut\Bstrut & \textbf{-0.77\%} &&	\textbf{-1.27\%} && \textbf{-2.25\%}  \Tstrut\Bstrut\\
\hline\hline
\textit{BangkokMarketVidevo} \textbf{(C)} & 0.08\% && -0.12\% && -0.49\% & -0.31\% && -0.58\% && -1.18\% \Tstrut\Bstrut\\
\textit{BuntingHangingAcrossHongKongVidevo} \textbf{(C)} & 0.05\% && -0.49\% && -1.02\% & -0.11\% && -0.38\% &&	-1.53\%  \Tstrut\Bstrut\\	
\textit{BusyHongKongStreetVidevo} \textbf{(C)}	& 0.08\% && 0.07\% && -0.10\% & -0.04\% && -0.31\% && -0.79\%  \Tstrut\Bstrut\\
\textit{CalmingWaterBVITexture} \textbf{(C)} & -0.07\% && 0.00\% && -0.07\% & -0.34\% && -0.36\% && -0.81\%  \Tstrut\Bstrut\\	
\hline
\textbf{non-CTC ClassC Overall} & \textbf{0.03\%} && \textbf{-0.14\%} && \textbf{-0.42\%} & \textbf{-0.20\%} && \textbf{-0.41\%} && \textbf{-1.08\%}  \Tstrut\Bstrut\\
\hline
\textit{SkyscraperBangkokVidevo} \textbf{(D)} & -0.23\% && -0.89\% && -1.00\% & -0.70\% && -1.53\% &&	-2.93\%\Tstrut\Bstrut\\	
\textit{VegetableMarketS2LIVENetFlix} \textbf{(D)} & -0.24\% && -0.59\% && -0.73\% & -0.54\% && -1.24\% && -2.28\%\Tstrut\Bstrut\\
\textit{VeniceSceneIRIS} \textbf{(D)} & -0.49\% && -1.32\% && -1.61\% & -0.87\% && -1.80\% &&	-2.64\%\Tstrut\Bstrut\\	
\textit{WalkingDownKhaoStreetVidevo} \textbf{(D)} & 0.09\% && -0.49\% && -1.11\% & -0.24\% && -0.82\% && -1.54\%\Tstrut\Bstrut\\
\hline
\textbf{non-CTC ClassD Overall} & \textbf{-0.22\%} && \textbf{-0.82\%} && \textbf{-1.11\%} & \textbf{-0.59\%} &&	\textbf{-1.35\%} && \textbf{-2.35\%}\Tstrut\Bstrut\\
\Xhline{3\arrayrulewidth}
\textbf{non-CTC ClassC\&D Overall} & \textbf{-0.09\%} && \textbf{-0.48\%}	&& \textbf{-0.77\%} \Tstrut\Bstrut & \textbf{-0.39\%} &&	\textbf{-0.88\%} && \textbf{-1.71\%}  \Tstrut\Bstrut\\
\hline
\hline

\end{tabular}
\end{table*}

\begin{table*}
\caption{CompetitionCNN coding complexity on CTC and non-CTC sequences\\for RA, LDB and LDP configurations, entire sequence.}
\label{tab:complex}
\centering
\begin{tabular}{c|cc|cc|cc} 
\hline
\hline
\multirow{2}{*}{Class} & \multicolumn{2}{c|}{RA} & \multicolumn{2}{c|}{LDB} & \multicolumn{2}{c}{LDP}\Tstrut\Bstrut\\
\cline{2-7}
& EncTime & DecTime  & EncTime & DecTime & EncTime & DecTime \Tstrut\Bstrut\\
\hline
CTC ClassC & 767\% & 290\% & 982\% & 231\% & 691\% & 221\% \Tstrut\Bstrut\\
\hline
CTC ClassD & 690\% & 264\% & 845\% & 220\% & 599\% & 204\% \Tstrut\Bstrut\\
\hline
non-CTC ClassC & 730\% & 220\% & 855\% & 180\% & 660\% & 190\% \Tstrut\Bstrut\\
\hline
non-CTC ClassD & 878\% & 281\% & 1171\% & 228\% & 783\% & 223\% \Tstrut\Bstrut\\
\hline
\hline
\end{tabular}
\end{table*}

\subsection{Comparison between three main architectures}
The ScratchCNN architecture, as defined in \cite{murn2020interpreting}, required different networks to be trained for each fractional shift. With the introduction of SharedCNN and CompetitionCNN, a single network can be trained to approximate all possible fractional shifts. This improvement allows the NN itself to share knowledge through the first two layers to decide how best to learn and define specific interpolation filters. With additional improvements in the training framework, these new architectures can lead to lower loss values when compared to ScratchCNN, and higher compression efficiency when integrated within the codec.

As presented in Table \ref{tab:multilinear}, when tested as switchable filter sets within VVC on 32 frames, both SharedCNN and CompetitionCNN exhibit higher coding gains compared to the previous ScratchCNN approach because of the inclusion of a two-layer trunk shared across all network branches. Another significant improvement introduced with the CompetitionCNN network is the competition of the branches amongst themselves when trained on a dataset, rather than being trained on subsets. As CompetitionCNN has the best coding performance among all approaches, with SharedCNN considered as an intermediate step, the learned filters extracted from it are subsequently tested on full non-CTC and CTC sequences, as presented in the next subsection.

\begin{figure*}[ht!]
    \centering
    \includegraphics[width=\textwidth]{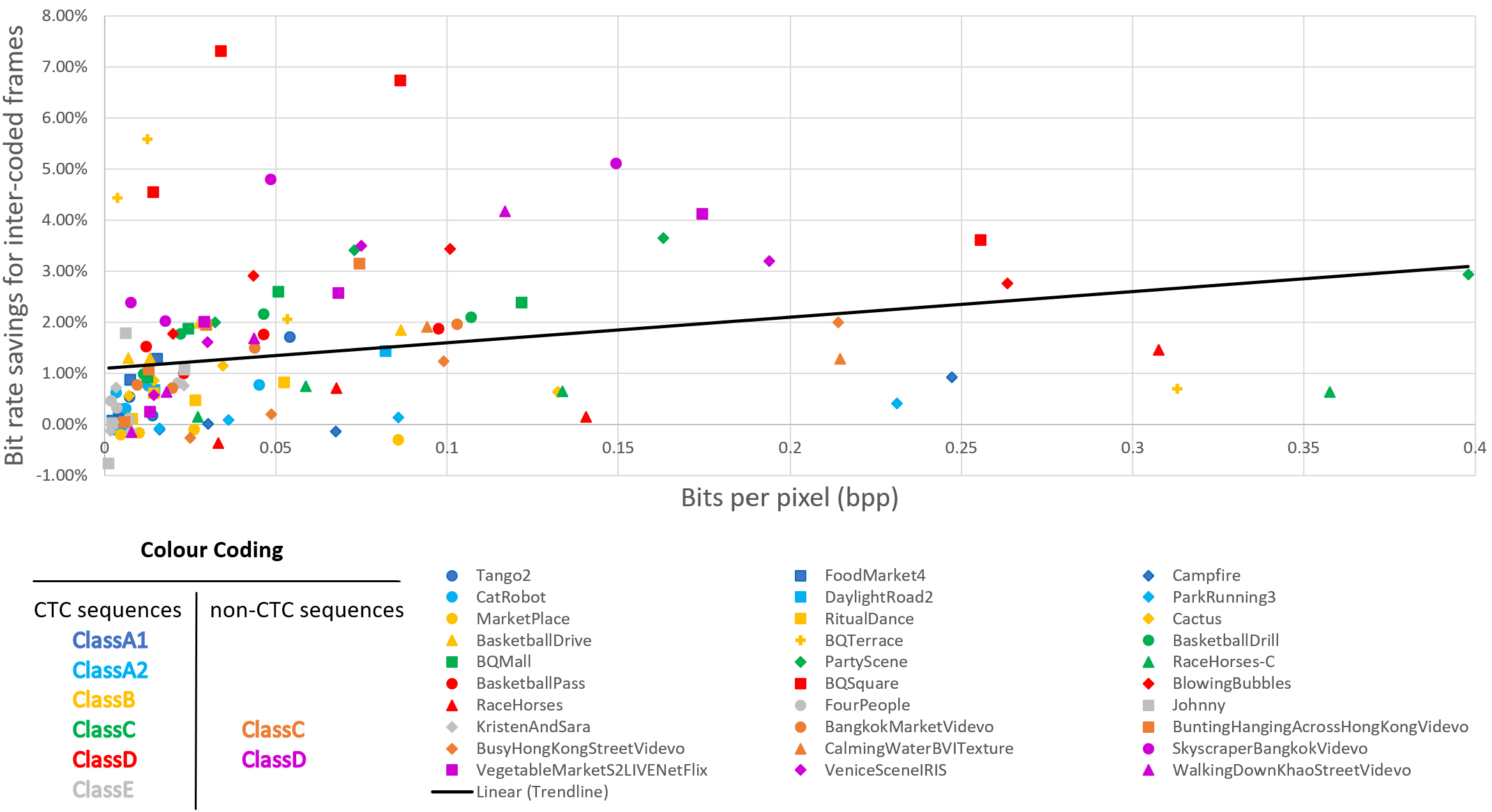}
    \caption{Bit rate savings of the CompetitionCNN switchable filter implementation in VVC for various video sequences at 4 reference bit rates - for QPs 22, 27, 32, 37. Higher QP indicates lower bit rate. The savings are outlined against the bits per pixel needed to encode the first $31$ inter-predicted frames of each sequence with the LDP configuration. A linear trendline is computed that takes into account every plot in the graph.}
    \label{fig:bpp}
\end{figure*}

\subsection{Compression performance evaluation}

A comparison between ScratchCNN and CompetitionCNN on Class C and D sequences for RA, LDB and LDP configuration is summarised in Table \ref{tab:bdr_configs}.

On average, coding gains have improved across all configurations and classes, from $-2.04\%$ to $-2.58\%$ for Class D, LDP, for example. Significant coding gains are featured for the $BQSquare$ sequence, a $-5.54\%$ improvement in compression efficiency compared to the baseline. The three-stage training framework has stabilised the performance of CompetitionCNN and allowed it to generalise well across varied video content, as CompetitionCNN displays coding gains for every sequence in every configuration. On the contrary, ScratchCNN performs inconsistently for some sequences like \textit{RaceHorses}, across both C and D classes. Additionally, CompetitionCNN demonstrates only minor coding efficiency differences for different classes in RA, unlike ScratchCNN filters, whose performance notably diminishes when tested on higher resolution sequences. 

As stated in Section \ref{sub:data}, the networks were trained on a Class D CTC sequence, \textit{BlowingBubbles}. In order to verify that both networks are not overfitting, tests on additional C and D sequences were performed, taken from the Bristol database for deep video compression \cite{ma2020bvidvc}. The database contains a total of $800$ video sequences, created from $200$ $4$K sequences, which were carefully selected from public video databases, such as Videvo Free Stock Video Footage \cite{videvo}, IRIS32 Free $4$K Footage \cite{iris}, BVI-Texture \cite{bvit}, and LIVE-Netflix \cite{bampis2018perceptually} among others. The $4$K sequences were spatially downsampled to resolutions of $1920\times1080$, $960\times540$ (Class C) and $480\times270$ (Class D) and truncated to 64 frames without scene cuts. 

For the experiments presented in this paper, $8$ video sequences were selected according to their low-level video features, spatial information (SI) and temporal information (TI). The features of these sequences accurately represent the general SI-TI plots of their respective class. CompetitionCNN retains its good predictive performance on these non-CTC sequences, with the compression efficiency reaching $-2.93\%$ for a Class D sequence in the LDP configuration. On the contrary, ScratchCNN achieves $-1.11\%$ coding gains on average for Class D with the LDP configuration, with this performance diminishing significantly when tested on Class C sequences.

Lastly, CompetitionCNN's coding complexity for CTC and non-CTC sequences is outlined in Table~\ref{tab:complex}. While prior approaches that had been proposed for HEVC required GPU coding environments \cite{ChromaMC,Yan2019,oneforall}, the proposed approach can be applied in conventional CPU-based codec environments, with tolerable complexity increases in both encoding and decoding time. In \cite{murn2020interpreting}, a full three-layer network was implemented within VVC and demonstrated increases in encoding complexity by up to $380$ times. The CompetitionCNN implementation that only requires $13\times13$ filters displays a $6$ to $11$ encoding time increase and a $2$ decoding time increase, when compared to the baseline. To allow fair comparison, Single Instruction/Multiple Data (SIMD) instructions were disabled for the VVC filters.

As explained in Section~\ref{subsec:learnedcoeffs}, there are possibilities for further complexity optimisations, most prominently in terms of integerisation of the learned coefficients or reducing the size of convolutional kernels, resulting in a smaller filter size.

\begin{figure*}
\centering
\subfloat[BasketballPass, 2nd frame, QP $27$, LDP configuration]{
    \includegraphics[width=0.45\textwidth]{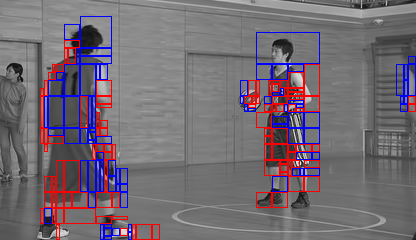}
    \label{fig:bbpass}
}
\hfill
\subfloat[BQSquare, 465th frame, QP $27$, LDP configuration]{
    \includegraphics[width=0.45\textwidth]{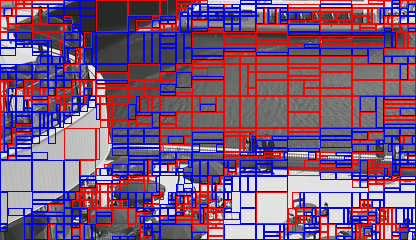}
    \label{fig:bqsquare}
}
\caption{Sub-pixel motion compensated blocks filter selection visualisation. Red CUs indicate the choice of CompetitionCNN filter set, blue CUs indicate the choice of the traditional filter set}
\label{fig:frameexamples}
\end{figure*}

\subsection{Evaluation on higher resolution video sequences}

A pattern emerges when examining the results presented in Table~\ref{tab:bdr_configs}. For both CTC and non-CTC sequences, higher coding gains are achieved for the lower resolution class. This occurrence can partly be explained with the training dataset, a Class D sequence. Therefore, additional experiments have been carried out. 

In Fig.~\ref{fig:bpp}, the bit rate savings of the proposed switchable filter approach are plotted against the bits per pixel (bpp) needed to encode the first $31$ inter-predicted frames of CTC and non-CTC video sequences, for 4 QPs (22, 27, 32, 37) under the LDP configuration. As opposed to Tables \ref{tab:multilinear} and \ref{tab:bdr_configs}, the savings are calculated only for inter-coded frames and positive values indicate coding gains. The described video quality evaluation methodology was first presented in \cite{vqevaluation}, with the script which computes this verification metric available in \cite{bbcinterpcurves}. In this paper, the methodology is extended by computing the bit rate savings at specific points defined by reference QP bit rates and against the bits per pixel.

A linear trendline suggests that higher bit rate savings are achieved for sequences which require more bits per pixel. The bit rate of inter-coded frames increases with more inter-predicted blocks and more motion vectors to be transmitted, meaning there are actually savings that can be achieved. It is more important to have methods that enhance compression of such sequences for which low bit rates are difficult to achieve. Video sequences that are easier to encode and for which the CompetitionCNN switchable filter set does not provide significant bit rate savings are concentrated at the bottom-left of the graph. These sequences generally belong to higher resolution classes - Class A1 ($4$K), Class A2 ($4$K), Class B ($1080$p) and Class E ($1080$p). Both CTC and non-CTC sequences of classes C and D display higher bit rate savings and usually require more bits to encode a pixel. $30$ out of the $40$ plot points with a bits per pixel value higher than $0.05$ belong to these two classes.

As the CompetitionCNN switchable filter implementation works better for sequences with higher bits per pixel values, and those usually correspond to lower resolution classes, we have provided full coding performance results for the selected classes for CTC and non-CTC sequences, shown in Table~\ref{tab:bdr_configs}.

\subsection{Learned interpolation filter selection analysis}

To explore why adding an interpolation filter set extracted from the learned CompetitionCNN brings coding gains, a filter selection ratio test was performed. Each CU within a frame-to-be-encoded can choose between the learned interpolation filter set and the traditional filter set during the Rate-distortion Optimisation (RDO) process. The ratio of CompetitionCNN-based filters being selected among all sub-pixel motion compensated CUs is presented in Table~\ref{tab:hitratio}. The results are presented for Class C and Class D CTC sequences, encoded with the LDP configuration. The hit ratio hovers around $50\%$, suggesting that the CompetitionCNN training framework, presented in Section~\ref{subsec:traincompcnn}, has ensured the development of true switchable filters, and not filters that could fully replace existing ones.

To visualise this selection process on inter-predicted frames, two examples are shown in Fig.~\ref{fig:frameexamples}. In Fig.~\ref{fig:bbpass}, the learned interpolation filter set is used on the outlines of the moving basketball players which are partitioned into smaller blocks. In Fig.~\ref{fig:bqsquare}, the learned filters are predominantly used on CUs positioned over the moving water, a notably challenging pattern to encode in visual data compression.

\begin{table}
\footnotesize
\caption{Selection ratio of CompetitionCNN filters among fractionally interpolated blocks, at different quantisation levels, LDP configuration}
\label{tab:hitratio}
\centering
\begin{tabular}{l|c|c|c|c} 
\hline
\hline
Class & QP 22 & QP 27 & QP 32 & QP 37 \Tstrut\Bstrut\\
\hline
Class C (CTC) & 47.20\% & 49.80\% & 47.55\% & 38.03\% \Tstrut\Bstrut\\
\hline
Class D (CTC) & 46.23\% & 50.25\% & 46.92\% & 38.87\% \Tstrut\Bstrut\\
\hline
\hline
\end{tabular}
\end{table}

\section{Conclusions}
A novel approach for developing sub-pixel interpolation filters using simplified, shared and interpreted CNNs was presented in this paper. A novel architecture containing two shared layers and a last layer with $15$ branches was introduced, alongside a new training framework, in order to maximise the coding performance of the newly formed interpolation filters. The linear structure of the network allows layers to collapse into a $2$D matrix for application during inference time. In this way, it is possible to directly extract and visualise the filter coefficients, i.e. to interpret what was learned and to further derive the relevant coefficients for the final pixel prediction within the filter set.

The approach was tested in the context of the challenging VVC codec as switchable interpolation filters. As shown in the experimental results, the new NN-based filters, extracted from a simple linear network, display considerable compression performance for lower resolution sequences while offering a much more sensible trade-off in increased coding complexity compared to prior approaches in this area, paving the way for possible practical implementations of NN-based interpolation filters within video encoders.

Further improvements and simplifications to the network training could be done, by adding an adaptive weight quantisation metric in the loss function, potentially resulting in optimised integerised representations of the learned filter coefficients. Finally, tackling the challenging problem of integrating the introduced method with all VVC tools would be beneficial, as it would allow for a verification of coding performance results under VTM CTC, along with a comparison of all prior approaches from related NN work that was proposed for HEVC.

\bibliographystyle{IEEEbib}
\bibliography{references}

\end{document}